\documentstyle[aps,eqsecnum]{revtex}
 
\begin{document}
\title{Second post-Newtonian radiative evolution of the relative orientations 
of angular momenta in spinning compact binaries}
\author{L\'{a}szl\'{o} \'{A}. Gergely}
\address{Laboratoire de Physique Th\'{e}orique, Universit\'{e} Louis Pasteur, 3-5 rue
de l'Universit\'{e} 67084 Strasbourg, France\\
and KFKI Research Institute for Particle and Nuclear Physics, Budapest 114,
P.O.Box 49, H-1525 Hungary}
\maketitle
 
\begin{abstract}
The radiative evolution of the relative orientations of the spin and orbital
angular momentum vectors ${\bf S}_{{\bf 1}}{\bf ,\ S}_{{\bf 2}}$ and ${\bf
L, }$ characterizing a binary system on eccentric orbit is studied up to the
second post-Newtonian order.
As an intermediate result, all Burke-Thorne type instantaneous radiative
changes in the spins are shown to average out over a radial period.
It is proved that spin-orbit and spin-spin terms contribute to the radiative
angular evolution equations, while Newtonian, first and second post-Newtonian
terms together with the leading order tail terms do not. In complement to the
spin-orbit contribution, given earlier,
the spin-spin contribution is computed and split into two-body and
self-interaction parts. The latter provide the second post-Newtonian order
corrections to the $3/2$ order Lense-Thirring description.
\end{abstract}
 
\section{Introduction}
 
Ongoing theoretical studies of the gravitational radiation emitted by
compact binaries and of the radiation reaction on the orbit are strongly
motivated by the desire to detect gravitational waves using the LIGO \cite
{LIGO} / VIRGO \cite{VIRGO} interferometers. In order to extract the signal
from the experimental data one needs templates of sufficient accuracy. Based
on the expectation that during the adiabatic regime the radiation reaction
will circularize the orbit \cite{Peters}, templates for circular orbits are
computed. If the orbits are eccentric, the detectability of the
gravitational waves by using circular templates decreased \cite{PoMa}.
Quinlan and Shapiro\cite{QuSha} and Hills and Bender\cite{HiBe} argue for a
significant number of eccentric binaries in galactic nuclei for which the
time before plunging is insufficient for circularization. A faithful
description of the behavior of these binaries in the adiabatic regime
requires not only high post-Newtonian orders, but also eccentric orbits.
Such a generic treatment, valid up to the second post-Newtonian order was
provided by Gopakumar and Iyer \cite{GI}.
 
If the spins of the neutron stars/black holes which form the binary are 
equally considered, the number of kinematical variables increases significantly 
and the dynamics complicates. Currently a second post-Newtonian order accurate
description of the motion of the spinning binary \cite{BOC}-\cite{Kidder} is
available. Spin-orbit and spin-spin type contributions appear at the $3/2$
and second post-Newtonian orders, respectively. There is one notable
exception over this rule: the spin-orbit and spin-spin terms in the spin
precession equations 
\begin{eqnarray}
{\bf \dot S_1}&=&{\frac{G}{c^2r^3}}
\left(\frac{4m_1+3m_2}{2m_1}{\bf L_N - S_2} +
\frac{3}{r^2}({\bf r\cdot S_2}){\bf r}\right) \times{\bf S_1}\ ,  \nonumber
\\
{\bf \dot S_2}&=&{\frac{G}{c^2r^3}}
\left(\frac{4m_2+3m_1}{2m_2}{\bf L_N - S_1
} + \frac{3}{r^2}({\bf r\cdot S_1}){\bf r}\right) \times{\bf S_2}\ \ 
\label{Sprec}
\end{eqnarray}
generate first and $3/2$ post-Newtonian order 
changes in the angles between the angular momenta vectors
\footnote{
We evaluate the post-Newtonian order of a correction $\delta f$ to some
quantity $f$ through the following prescription. Count the powers of $
\epsilon \approx Gm/rc^{2}\approx v^{2}/c^{2}$, representing one
post-Newtonian order, in both $\delta f$ and $f$. ($G$ is the gravitational
constant, $c$ the speed of light, $m$ the mass of binary, $r$ and $v$ the
orbital radius and velocity; the characteristic radius for a compact object
being $R\approx Gm/c^{2}.$) The post-Newtonian order is their ratio $PN{\cal
O}(\delta f)={\cal O}_{{\cal \epsilon }}(\delta f)/{\cal O}_{{\cal \epsilon }
}(f)$. When the time derivative $\dot{f}$ is given, we can evaluate the
order of the change $\delta f$ generated during a characteristic
time-interval of the quasi-periodic motion: $PN{\cal O}(\delta f)={\cal O}_{
{\cal \epsilon }}(\dot{f}){\cal O}_{{\cal \epsilon }}(T)/{\cal O}_{{\cal
\epsilon }}(f)$. Here $T$ is the radial period and ${\cal O}_{{\cal \epsilon
}}(T)=r\epsilon ^{-1/2}/c$. For a dimensionless quantity, typically an angle
variable $\alpha $, the simple estimate $PN{\cal O}(\delta \alpha )={\cal O}
_{{\cal \epsilon }}(\dot{\alpha}){\cal O}_{{\cal \epsilon }}(T)$ holds.
\par
A numerical example shows that for a binary consisting of massive black
holes with mass $100M_{\odot }$ and orbital radius of $10^{4}$ km, {\em i.e.}
, one-hundredth of the Hulse-Taylor system, the post-Newtonian parameter is $
\epsilon \approx Gm/rc^{2}\approx 10^{-2}$.
\par
We also assume rapid rotation, thus the characteristic velocity of rotation $V$
is commesurable with the velocity of light ${\cal O}(V/c)=1$. If this assumption 
is lifted, the spin-spin effects appear at higher post-Newtonian orders 
than is assumed here.
}.
Consequences of the spin precession equations (\ref{Sprec}) and of the conservation 
of the total angular momentum ${\bf J}={\bf L}+{\bf S_1}+{\bf S_2}$ holding in the 
second post-Newtonian order are summarized in Table 1.
In the present paper the angular momenta, their directions, 
magnitudes and the angles, all enlisted in Table 1 will be needed to a precision at 
which precessional effects do not contribute.

All spin terms in the radiation reaction up to the $3/2$ post-Newtonian
order were evaluated in \cite{GPV3}, starting from the expressions of the
radiated power and total angular momentum loss derived by Kidder \cite
{Kidder} and from the Burke-Thorne potential \cite{BT}. The employed
averaging method, based on a suitably introduced radial parameter $\chi $
and on the residue theorem was described in detail in \cite{param}.
 
The second-order spin contribution to the radiation reaction was considered
recently in \cite{spinspin}. There we have developed the toolchest for
computing secular radiative effects of spin-spin origin, essentially by
proposing a description in terms of constants of the motion and the angular
average $\bar{L}$ of the magnitude of the orbital angular momentum
$L(\chi )$.
The latter depends on the Keplerian true anomaly $\chi $ through a
spin-spin term:
\begin{eqnarray}
L(\chi ) &=&\bar{L}-\frac{G\mu ^{2}}{2c^{2}\bar{L}^{3}}S_{1}S_{2}\sin
\!\kappa _{1}\sin \!\kappa _{2}\{2\bar{A}\cos [\chi \!+2(\!\psi _{0}\!-\!
\bar{\psi})]  \nonumber \\
&&+(3Gm\mu \!+\!2\bar{A}\cos \!\chi )\cos 2(\chi \!+\!\psi _{0}\!-\!\bar{
\psi })\}\ .  \label{LchiAng1}
\end{eqnarray}
The quantity $\bar{A}=(G^{2}m^{2}\mu ^{2}+2E\bar{L}^{2}/\mu )^{1/2}$ is 
the magnitude of the Laplace-Runge-Lenz vector for a Keplerian motion 
characterized by the energy $E$ and magnitude of orbital angular momentum 
$\bar{L}$. 
The angles $\kappa _{i}=\cos ^{-1}({\bf \hat{S}_{i}\cdot \hat{L})},\ (i=1,2)$
together with $\gamma =\cos ^{-1}({\bf \hat{S}_{1}\cdot \hat{S}_{2}).}$ 
characterize the relative orientation of the angular momentum vectors 
${\bf S}_{{\bf 1}}{\bf ,\ S}_{{\bf 2}}$ and ${\bf L}$. 

This description enabled us to introduce the generalized true anomaly
parametrization $r(\chi )$ for the radial motion:
(Eq. (3.7) of Ref.\cite{spinspin}).
By employing $\chi$ as radial variable, the average of 
a generic function $f$ over a radial period $T$ \begin{equation}
\langle f\rangle =\frac{1}{T}
\int_{0}^{2\pi }f(\chi )\frac{1}{\dot r}\frac{dr}{d\chi}d\chi
\   \label{ave}
\end{equation}
is computed by Taylor expanding $\dot r^2$ taken from the radial equation, 
by substituting  
\begin{equation}
\frac{dr}{d\chi }=\frac{1}{2}\left( \frac{1}{r_{min}}-\frac{1}{r_{max}}
\right) r^{2}\sin \chi \   \label{drdchi}
\end{equation}
($r_{{}_{{}_{min}^{max}}}$ being the turning points)
and finally, by switching to the complex parameter $z=\exp (i\chi )$. Then the 
residue theorem is employed in the computation of the integral (\ref{ave}), 
which in the 
majority of cases is given by the residues in the origin \cite{param}.
 
As application, we have
computed the secular losses of the energy $E$ and of $\bar{L}$. They depend
on $E,$ $\bar{L},$ on the angles$\ \psi _{i}$ subtended by the node line
with the projections of the spins in the plane of the orbit and on the
angles $\kappa _{i}$ and $\gamma $. The enlisted
five angles are constrained by two algebraic relations \cite{GPV3}:
 
\begin{eqnarray}
0 &=&S_{1}\sin \kappa _{1}\cos \psi _{1}+S_{2}\sin \kappa _{2}\cos \psi _{2}
\label{ang1} \\
\cos \gamma &=&\cos \kappa _{1}\cos \kappa _{2}+\sin \kappa _{1}\sin \kappa
_{2}\cos \Delta \psi \ ,  \label{ang2}
\end{eqnarray}
where $\Delta \psi =\psi _{2}-\psi _{1}$. We also introduce here the notation $
\bar{\psi}=(\psi _{1}+\psi _{2})/2.$ In order to have a closed system of
differential equations, beside $dE/dt$ and $dL/dt$ computed in \cite
{spinspin} the radiative change of the angles $\kappa _{i}$ and $\gamma $ is
required. This was given to $3/2$ post-Newtonian order accuracy in \cite
{GPV3}. It is the purpose of the present paper to compute the radiative
evolution of the angles $\kappa _{i}$ and $\gamma $ up to the second
post-Newtonian order by the inclusion of the spin-spin terms.

\begin{table}[tbp]
\caption{The effect of precessions on angular momenta.}
\begin{tabular}{cccc}
quantity $(f)$       & leading order change ${\cal O}(\delta f)$ & 
origin of the leading order change & constant at order  
\\
\tableline 
${\bf S_i}\ ,\ {\bf \hat S_i}={\bf S_i}/S_i\ ,\ 
\gamma =\cos ^{-1}({\bf \hat{S}_{1}\cdot \hat{S}_{2})}$ & 
$\epsilon$ & spin-orbit &
$\epsilon^{1/2}$  
\\
${\bf L}\ ,\ {\bf \hat L}={\bf L}/L$ & 
$\epsilon^{3/2}$ & spin-orbit &
$\epsilon$  
\\
$\kappa _{i}=\cos ^{-1}({\bf \hat{S}_{i}\cdot \hat{L})}$ & 
$\epsilon^{3/2}$ & spin-spin &
$\epsilon$  
\\
$L$ & 
$\epsilon^2$ & spin-spin &
$\epsilon^{3/2}$  
\\
$S_i$ & 
$\ge\epsilon^2 $ & higher order effects &
$\epsilon^{3/2}$  
\end{tabular}
\end{table}
  
We complete the task by passing through the following steps. A well-known
result states \cite{MTW} that no leading-order gravitational radiation
leaves an axisymmetric body, as its quadrupole moment is a constant. It is
also known that the radiative loss $d{\bf S}_{{\bf i}}/dt$ in the spins of
the axisymmetric neutron stars or black holes forming a binary is two
post-Newtonian orders higher that the leading radiative loss $d{\bf L/}dt$
of the orbital angular momentum \cite{ACST}. In Sec. 2 we derive a related
generic result, which holds for binaries with axisymmetric constituents. We
show that axisymmetry implies that the leading order {\it secular}
contribution to the radiative change of the spins (which would appear at the
second post-Newtonian order after the leading radiation terms) averages out
during one cycle of the quasi-periodic motion. This result simplifies the
study of the secular evolution of the angles. Indeed we will drop all terms
of the form $d{\bf S_{i}}/dt$ multiplied with any quantity which is a
constant to the required accuracy.

The rest of the paper is organized as follows. In Sec. 3 we discuss the
radiative change in the angles $\kappa _{i}$ and $\gamma $ by considering
the Newtonian ($N$), first post-Newtonian ($PN$), second post-Newtonian ($2PN
$), spin-orbit ($SO$), spin-spin ($SS$) and leading order tail
contributions.
We show that the $N,\ PN,\ 2PN$ and tail terms drop out as a consequence of
the property of the respective parts in $d{\bf J}/dt$ of being aligned with
the Newtonian part of the orbital angular momentum. The $SO$ terms were
given previously in \cite{GPV3}. We complete the list by computing in Sec. 4
the averaged $SS$ contributions. They contain
both two-body and self-interaction terms.
 
As a result we obtain the system of equations which govern the radiative
secular evolution of the relative orientations of the angular momentum
vectors ${\bf S}_{{\bf 1}}{\bf ,\ S}_{{\bf 2}}$ and ${\bf L}$ up to the
second post-Newtonian order.
 
\section{Radiative loss in the spins}
 
In \cite{GPV2} we have derived the radiative loss in the spin ${\bf S_{i}}$
of the $i^{th}$ axisymmetric body, following \cite{ACST}. The integral over
the volume of the body of the moment of the reaction force (the sign swapped
gradient of the Burke-Thorne potential) gave:
\begin{equation}
\frac{d\left( {\bf S_{i}}\right) _{\mu }}{dt}={\frac{2G}{
5c^{5}\Omega _{i}}}\left( {\frac{\Theta _{i}}{\Theta _{i}^{\prime }}}
-1\right) \epsilon _{\mu \nu \rho }I_{N}^{(5)\nu \sigma }S_{i}({\bf \hat{S}
_{i}})_{\rho }({\bf \hat{S}_{i}})_{\sigma }\ .  \label{Sdirdot}
\end{equation}
(There is no summation over $i.$) Each body is characterized by its principal
moments of inertia $\Theta _{i}$ and $\Theta _{i}^{\prime }$ and by the
angular velocity $\Omega _{i}=S_{i}/\Theta _{i}^{\prime }$, while $
I_{N}^{(5)\nu \sigma }$ is the $5^{th\text{ }}$time derivative of the
{\it system}'s Newtonian symmetric trace-free mass quadrupole moment. As the loss
in the spin vector described by Eq. (\ref{Sdirdot}) is of second
post-Newtonian order (above the leading radiation term in the orbital
angular momentum loss), Eq. (\ref{Sdirdot}) is still valid to this order for
{\it approximately} axisymmetric bodies, with the deviation from axisymmetry
being of any post-Newtonian order. We emphasize that Eq. (\ref{Sdirdot})
implies $d{\bf S_{i}}/dt=$ $S_{i}d{\bf \hat{S}_{i}}/dt$, therefore to this order 
the radiation reaction will change the orientation but not the magnitude of the 
spin vectors. 
We also stress that $d{\bf \hat{S}_{i}}/dt$ generates $3/2$ post-Newtonian order 
radiative changes $\delta_{rad} {\bf \hat{S}_{i}}$ above the leading order 
radiation losses (in $L$ for example).
 
In coordinates $(x,y,z)=r(\cos \psi ,\sin \psi ,0)$ the spins are expressed
as ${\bf S_{i}=S}_{i}(\sin \kappa _{i}\cos \psi _{i},\sin \kappa _{i}\sin
\psi _{i},\cos \kappa _{i}).$ By employing the Keplerian equation of motion $
{\bf a}_{N}=-Gm{\bf r/}r^{3}$ and the radial equation $\dot{r}^{2}=2E/\mu
+2Gm/r-\bar{L}^{2}/\mu ^{2}r^{2}$ in Eq. (\ref{Sdirdot}), we obtain the
instantaneous spin-loss equation:
\begin{eqnarray}
\frac{1}{S_{i}}\frac{d\left( {\bf S_{i}}\right) _{\mu }}{dt} &=&{\frac{
2G^{2}m\sin \kappa _{i}}{5c^{5}\mu ^{2}r^{7}\Omega _{i}}}\left( {\frac{
\Theta _{i}}{\Theta _{i}^{\prime }}}-1\right) \Lambda _{i\mu }  \nonumber \\
\Lambda _{i1} &=&\cos \kappa _{i}[a_{1}\sin (2\psi -\psi _{i})+a_{2}\cos
(2\psi -\psi _{i})+a_{3}\sin \psi _{i}]\   \nonumber \\
\Lambda _{i2} &=&\cos \kappa _{i}[a_{2}\sin (2\psi -\psi _{i})-a_{1}\cos
(2\psi -\psi _{i})-a_{3}\cos \psi _{i}]  \nonumber \\
\Lambda _{i3} &=&-\sin \kappa _{i}[a_{1}\sin 2(\psi -\psi _{i})+a_{2}\cos
2(\psi -\psi _{i})]\ ,  \label{Sdirlossinst}
\end{eqnarray}
with the coefficients
\begin{eqnarray}
a_{1} &=&\mu r\dot{r}(12\mu Er^{2}+20Gm\mu ^{2}r+45\bar{L}^{2})  \nonumber \\
a_{2} &=&4\bar{L}(18\mu Er^{2}+20Gm\mu ^{2}r-15\bar{L}^{2})  \nonumber \\
a_{3} &=&\mu r\dot{r}(12\mu Er^{2}+20Gm\mu ^{2}r-15\bar{L}^{2})\ .
\end{eqnarray}
 
In order to obtain the secular radiative changes of the spins we insert the
Newtonian expressions
\begin{equation}
\dot{r}=\frac{\bar{A}}{\bar{L}}\sin \chi \ ,\qquad \ r=\frac{\bar{L}^{2}}{
\mu (Gm\mu +\bar{A}\cos \chi )}\ ,\qquad \psi =\chi +\psi _{0}\ .
\label{Newt}
\end{equation}
in Eq. (\ref{Sdirlossinst}). 
 
Then we average by the method described in the Introduction 
(see also \cite{spinspin}) employing the Newtonian expression
$(1/{\dot r})(dr/d\chi)=dt/d\chi =\mu r^{2}/\bar{L}.$ 
We obtain the generic result that {\it the
radiative change in the spins averages to zero in the second post-Newtonian
approximation}:
\begin{equation}
\left\langle \frac{d{\bf S_{i}}}{dt}\right\rangle =0\ .  \label{nospinave}
\end{equation}
Therefore the averaging-out property of some particular projections of the
spin losses found in \cite{GPV3} and \cite{GPV2} also emerge.
 
As the first correction to both the acceleration and the Burke-Thorne
potential is one post-Newtonian order higher than the respective leading
terms \cite{Blanchet}, the averaging-out property of the radiative change in
the spins derived above holds at $5/2$ post-Newtonian orders as well.
 
\section{Radiative evolution of the angles $\kappa _{i}$ and $\gamma $}
 
We obtain the equations for the radiative evolution of $\kappa _{i}=\cos
^{-1}({\bf \hat{S}_{i}\cdot \hat{L})},\ (i=1,2)$ and $\gamma =\cos ^{-1}(
{\bf \hat{S}_{1}\cdot \hat{S}_{2})}$ by differentiating the defining
relations of these angles.
 
The simplest is the evolution of the angle $\gamma $ spanned by the two spin
vectors:
 
\begin{equation}
\frac{d}{dt}\cos \gamma =\frac{d}{dt}\left( {\bf \hat{S}_{1}\cdot \hat{S}}_{
{\bf 2}}\right) ={\bf \hat{S}}_{{\bf 1}}{\bf \cdot }\frac{d{\bf \hat{S}}_{
{\bf 2}}}{dt}+{\bf \hat{S}}_{{\bf 2}}{\bf \cdot }\frac{d{\bf \hat{S}_{1}}}{
dt }\ .
\end{equation}
The terms of the right hand side were already evaluated in the framework of
the spin-orbit contributions in \cite{GPV3}. Though they do not contain the
spin explicitly, the angular velocities $\Omega _{i}$ are present in their
expression, Eq. (3.13) of \cite{GPV3}. An order of magnitude estimate shows
that the ratio of the usual spin-orbit terms (taken for example from the
losses of $\kappa _{i}$, Eqs. (3.10)-(3.12) of \cite{GPV3}) and such terms
is of the order unity for rapidly rotating objects. It was
computed in \cite{GPV3}, and one can check by simple inspection of Eq. (\ref
{nospinave}) that $\gamma $ will receive {\it no secular radiative change}.
 
The angles spanned by the spins with the orbital angular momentum evolve in
a more complicated fashion:
\begin{equation}
\frac{d}{dt}\cos \kappa _{i}=\frac{d}{dt}\left( {\bf \hat{S}_{i}\cdot }\frac{
{\bf L}}{L(\chi )}\right) =\frac{1}{L(\chi )}\left[ {\bf \hat{S}}_{{\bf i}}
{\bf \cdot }\frac{d{\bf (J-S}_{{\bf 1}}-{\bf S}_{{\bf 2}})}{dt}-\frac{dL}{dt}
\cos \kappa _{i}+{\bf L\cdot }\frac{d{\bf \hat{S}_{i}}}{dt}\right] \ .
\label{kappaloss1}
\end{equation}
In the forthcoming expressions we replace $L(\chi )\rightarrow \bar{L}$ in
all post-Newtonian terms. This is possible in the required second order
accuracy as $L(\chi )$ differs from the constant $\bar{L}$ in spin-spin
terms (see Eq. (\ref{LchiAng1})).
 
By inserting the following expression for the loss of the magnitude of
orbital angular momentum
\begin{equation}
\frac{dL}{dt}={\bf \hat{L}\cdot }\frac{d{\bf L}}{dt}={\bf \hat{L}\cdot }
\frac{d{\bf J}}{dt}-{\bf \hat{L}\cdot }\frac{d{\bf S_{1}}}{dt}-{\bf \hat{L}
\cdot }\frac{d{\bf S_{2}}}{dt}  \label{Lloss}
\end{equation}
into Eq. (\ref{kappaloss1}), and from ${\bf \hat{S}}_{{\bf i}}\cdot d{\bf S}
_{{\bf i}}/dt=0$ (no summation) we obtain:
\begin{equation}
\frac{d}{dt}\cos \kappa _{i}=\frac{1}{L(\chi )}({\bf \hat{S}}_{{\bf i}}-{\bf
\hat{L}}\cos \kappa _{i}){\bf \cdot }\frac{d{\bf J}}{dt}+\frac{1}{\bar{L}}
\left[ -{\bf \hat{S}}_{{\bf i}}{\bf \cdot }\frac{d{\bf S}_{{\bf j}}}{dt}+
{\bf \hat{L}\cdot }\left( \frac{d{\bf S_{1}}}{dt}+\frac{d{\bf S_{2}}}{dt}
\right) \cos \kappa _{i}+{\bf L\cdot }\frac{d{\bf \hat{S}_{i}}}{dt}\right] \
,  \label{kappaloss2}
\end{equation}
where $j\neq i.$ Remarkably all terms in the square bracket average out due
to Eq.(\ref{nospinave}). Therefore we will not consider them further.
 
For the evaluation of the first term of Eq. (\ref{kappaloss2}) we need the
loss in the total angular momentum $d{\bf J}/dt$. It has the following
structure\footnote{
To avoid confusion in notation, we denote the post-Newtonian, $2^{d}$
post-Newtonian, spin-orbit and spin-spin terms from a decomposition of a
quantity expressed in terms of ${\bf r,v,S}_{{\bf i}}$ and $\dot{r}$ by
lower-case indices $\ pn,\ 2pn,\ so$ and $ss$, respectively. This notation
applies to all terms of $d{\bf J}/dt$ computed in \cite{Kidder} and \cite{GI}
. After inserting the expressions for $v^{2}$, $\dot{r}^{2}$ and $\dot{r}$,
Eqs. (2.32), (2.33) of \cite{spinspin} and (\ref{Newt}), the terms of the
new decomposition will be denoted by the respective capital letters.} \cite
{Kidder} :
\begin{equation}
\frac{d{\bf J}}{dt}=\left( \frac{d{\bf J}}{dt}\right) _{N}+\left( \frac{d
{\bf J}}{dt}\right) _{pn}+\left( \frac{d{\bf J}}{dt}\right) _{2pn}+\left(
\frac{d{\bf J}}{dt}\right) _{so}+\left( \frac{d{\bf J}}{dt}\right)
_{ss}+\left( \frac{d{\bf J}}{dt}\right) _{tail}\ .  \label{Jloss}
\end{equation}
The leading order contribution was computed long time ago by Peters \cite
{Peters}, the $pn$ contribution by Junker and Sch\={a}fer \cite{JS}, the $so$
contribution by Kidder \cite{Kidder}, and the $2pn$ contribution by
Gopakumar and Iyer \cite{GI} in terms of ${\bf r,v,S}_{{\bf i}}$ and $\dot{r}
$. The method of computing the $ss$ contribution was also indicated in \cite
{Kidder}. The leading tail contribution appears at $3/2$ post-Newtonian
order and its time average was given by Rieth and Sch\"{a}fer \cite{RS} in
the form of a Fourier series.
 
A property we want to stress is the following:
\begin{equation}
\left( \frac{d{\bf J}}{dt}\right) _{N,\ pn,\ 2pn}=\Gamma _{0,\ 1,\ 2}\cdot
{\bf L}_{N}
\end{equation}
where the coefficients $\Gamma _{0,\ 1,\ 2}$ can be read from Eq. (3.28.a,b)
of \cite{Kidder} and from (3.9.d) of \cite{GI}. The same property applies to
the averaged tail contribution, given by Eq. (84) of \cite{RS}. To see
this we note that all coefficients
(denoted there by $_{n}\langle{\cal S}_{0}\rangle $)
of the spectral decomposition of the angular momentum loss
(denoted there by $\langle {\cal S}\rangle $) have only
$z-$components, thus:
\begin{equation}
\left\langle \frac{d{\bf J}}{dt}\right\rangle _{tail}=\Gamma _{tail}\cdot
{\bf L}_{N}\ .
\end{equation}
The next essential remark is that from
\begin{equation}
{\bf L=L}_{N}+{\bf L}_{PN}+{\bf L}_{2PN}+{\bf L}_{SO}=(1+\gamma _{1}+\gamma
_{2}){\bf L}_{N}+{\bf L}_{SO}
\end{equation}
we can express the orbital angular momentum as
\begin{equation}
{\bf L}_{N}=(1-\gamma _{1}-\gamma _{2}+\gamma _{1}^{2}){\bf L}-{\bf L}_{SO}\
.  \label{LN}
\end{equation}
Here the coefficients $\gamma _{1,\ 2}$ are given by Eqs. (2.9.b,d) of \cite
{Kidder} and they are of first and second post-Newtonian order,
respectively. Therefore up to an accuracy of second post-Newtonian order we
can write
\begin{equation}
\left( \frac{d{\bf J}}{dt}\right) _{N+\ pn\ +\ 2pn} =(\Gamma _{0}{\bf +}
\eta _{1}+\eta _{2}){\bf L-}\Gamma _{0}\cdot {\bf L}_{SO}\
\label{JlossNPN2PN}
\end{equation}
and
\begin{equation}
\left\langle \frac{d{\bf J}}{dt}\right\rangle _{tail} =\Gamma _{tail}\cdot
{\bf L}\ ,  \label{Jlosstailave}
\end{equation}
where the coefficients of the expansion (\ref{JlossNPN2PN}) are
given by \begin{eqnarray}
\eta _{1} &=&\Gamma _{1}-\Gamma _{0}\gamma _{1}  \nonumber \\
\eta _{2} &=&\Gamma _{2}-\Gamma _{1}\gamma _{1}-\Gamma _{0}(\gamma
_{2}-\gamma _{1}^{2})\ .
\end{eqnarray}
We will not need their explicit expressions. By inserting Eq. (\ref{JlossNPN2PN}) 
in Eq. (\ref{Jloss}), then the resulting expression in Eq. (\ref{kappaloss2}), 
due to the remark
\begin{equation}
({\bf \hat{S}}_{{\bf i}}-{\bf \hat{L}}\cos \kappa _{i})\cdot {\bf L=}0
\label{propr}
\end{equation}
we obtain for the first term of Eq. (\ref{kappaloss2}):
\begin{equation}
\frac{1}{L(\chi )}({\bf \hat{S}}_{{\bf i}}-{\bf \hat{L}}\cos \kappa
_{i})\cdot \left( \frac{d{\bf J}}{dt}\right) =\frac{1}{\bar{L}}({\bf \hat{S}}
_{{\bf i}}-{\bf \hat{L}}\cos \kappa _{i})\cdot \left[ \left( \frac{d{\bf J}}{
dt}\right) _{tail}+\left( \frac{d{\bf J}}{dt}\right) _{so}-\Gamma _{0}{\bf L}
_{SO}+\left( \frac{d{\bf J}}{dt}\right) _{ss}\right] \ .  \label{kappaloss3}
\end{equation}
The tail term averages out due to Eq. (\ref{Jlosstailave}) and property
(\ref{propr}).
 
As suggested by the notation, the second and third terms of Eq.
(\ref {kappaloss3}) are those spin-orbit contributions to the radiative loss of
the angles $\kappa _{i}$, which do not originate in the Burke-Thorne
potential. Indeed, modulo Burke-Thorne type terms they can be put into the
concise form (${\bf \hat{S}}_{{\bf i}}\cdot d{\bf \hat{L}/}dt)_{SO}$ which
was computed previously (Eqs. (3.8)-(3.10) of \cite{GPV3}). The averaged
expression for the spin-orbit type loss in $\kappa _{1}$ was also given by
Eq. (4.4) of \cite{GPV3}, and a similar expression can be found for the loss
of $\kappa _{2}$ by interchanging the indices $1\leftrightarrow 2$ and the
ratios $\eta =m_{2}/m_{1}\leftrightarrow \eta ^{-1}$.
 
The last term of Eq. (\ref{kappaloss3}), modulo Burke-Thorne type terms, is
the spin-spin part of the radiative evolution of the angles $\kappa _{i}$:
\begin{equation}
\left( \frac{d}{dt}\cos \kappa _{i}\right) _{SS}\simeq \frac{1}{\bar{L}}(
{\bf \hat{S}}_{{\bf i}}-{\bf \hat{L}}\cos \kappa _{i})\cdot \left( \frac{d
{\bf J}}{dt}\right) _{ss}\ ,  \label{kappaloss4}
\end{equation}
which will be computed in the next section. (We have denoted by $\simeq $
the equality modulo Burke-Thorne type terms.)
 
\section{Secular radiative evolution equations}
 
We start from the expression of $(d{\bf J}/dt)_{ss}$ given by Kidder \cite
{Kidder} in terms of the time derivatives of the mass quadrupole and
velocity quadrupole moments (see also Eq. (4.18) of \cite{spinspin} for the
respective expression with $c\neq 1\neq G$). The required $SO$ part of the
velocity quadrupole moment was computed first by Kidder \cite{Kidder} and
later verified by several authors. (Rieth and Sch\={a}fer \cite{RS} presented
a derivation based on a different spin supplementary condition while Owen,
Tagoshi and Ohasha \cite{OTT} have employed a $\delta$-function type
energy-momentum tensor.)
 
After computing the expression $(d{\bf J}/dt)_{ss}$ in detail, we rewrite
the last term of Eq. (\ref{kappaloss3}) as function of the radial variables $
r(\chi )$ and $\chi .$ The procedure we follow was described in detail in
\cite{spinspin}. As a result both self-interaction and two-body spin-spin
terms emerge:
\begin{eqnarray}
\left[ \frac{1}{\bar{L}}({\bf \hat{S}}_{{\bf i}}-{\bf \hat{L}}\cos \kappa
_{i})\cdot \left( \frac{d{\bf J}}{dt}\right) _{ss}\right] _{SS-self} &=&
\frac{2G^{3}m^{2}\mu \sin \kappa _{i}}{5c^{7}r^{6}}\left[ \left( \frac{S_{i}
}{m_{i}}\right) ^{2}\sin \kappa _{i}\cos \kappa _{i}+\left( \frac{S_{j}}{
m_{j}}\right) ^{2}\sin \kappa _{j}\cos \kappa _{j}\cos \Delta \psi \right] \\
\left[ \frac{1}{\bar{L}}({\bf \hat{S}}_{{\bf i}}-{\bf \hat{L}}\cos \kappa
_{i})\cdot \left( \frac{d{\bf J}}{dt}\right) _{ss}\right] _{S_{1}S_{2}} &=&
\frac{2G^{2}S_{1}S_{2}\sin \kappa _{i}}{5c^{7}\mu ^{2}\bar{L}^{2}r^{7}}\Bigl
\{u_{1}[\sin \kappa _{i}\cos \kappa _{j}+\sin \kappa _{j}\cos \kappa
_{i}\cos \Delta \psi ]  \nonumber \\
&&+u_{2}[\sin \kappa _{i}\cos \kappa _{j}\cos 2(\chi +\psi _{0}-\psi
_{i})+\sin \kappa _{j}\cos \kappa _{i}\cos 2(\chi +\psi _{0}-\bar{\psi})]
\nonumber \\
&&+\sin \chi \{u_{3}\sin \kappa _{j}\cos \kappa _{i}\sin (\psi _{j}-\psi
_{i})  \nonumber \\
&&+u_{4}[\sin \kappa _{i}\cos \kappa _{j}\sin 2(\chi +\psi _{0}-\psi
_{i})+\sin \kappa _{j}\cos \kappa _{i}\sin 2(\chi +\psi _{0}-\bar{\psi})]\}
\Bigr\}\ .
\end{eqnarray}
Here $j\neq i$ and the coefficients are
\begin{eqnarray}
u_{1} &=&-\bar{L}^{2}(12\mu Er^{2}+4Gm\mu ^{2}r-15\bar{L}^{2})  \nonumber \\
u_{2} &=&3\bar{L}^{2}(Gm\mu ^{2}r+3\bar{L}^{2})  \nonumber \\
u_{3} &=&3\mu \bar{A}r(-2\mu Er^{2}+5\bar{L}^{2})  \nonumber \\
u_{4} &=&3\mu \bar{A}r(-2\mu Er^{2}+3\bar{L}^{2})
\end{eqnarray}
 
The averaging procedure based on the parametrization $r(\chi )$ and on the
residue theorem yields the self-interaction and two-body spin-spin terms in
the secular radiative loss of the angles $\kappa _{i}$ and $\gamma $. We
enlist them together with the spin-orbit contributions:
\begin{eqnarray}
\left\langle \frac{d\gamma }{dt}\right\rangle &=&0  \label{gammaloss} \\
\left\langle \frac{d\kappa _{i}}{dt}\right\rangle &=&\left\langle \frac{
d\kappa _{i}}{dt}\right\rangle _{SO}+\left\langle \frac{d\kappa _{i}}{dt}
\right\rangle _{SS-self}+\left\langle \frac{d\kappa _{i}}{dt}\right\rangle
_{SS}  \label{kappaloss} \\
\left\langle \frac{d\kappa _{i}}{dt}\right\rangle _{SO} &&\ given\ by\ Eq.\
(4.4)\ of\ \cite{GPV3}  \label{kappalossSO} \\
\left\langle \frac{d\kappa _{i}}{dt}\right\rangle _{SS-self} &=&-\ \frac{
G^{2}m\mu (-2\mu E)^{3/2}}{20c^{7}\bar{L}^{9}}V_{1}\Biggl [\left( \frac{
S_{i} }{m_{i}}\right) ^{2}\sin \kappa _{i}\cos \kappa _{i}+\left( \frac{S_{j}
}{ m_{j}}\right) ^{2}\sin \kappa _{j}\cos \kappa _{j}\cos \Delta \psi \Biggr
]  \label{kappalossself} \\
\left\langle \frac{d\kappa _{i}}{dt}\right\rangle _{S_{1}S_{2}} &=&-\ \frac{
G^{2}(-2\mu E)^{3/2}S_{1}S_{2}}{20c^{7}\bar{L}^{9}\sin \kappa _{i}}\Bigl
\{V_{2}(\sin \kappa _{i}\cos \kappa _{j}+\sin \kappa _{j}\cos \kappa
_{i}\cos \Delta \psi )  \nonumber \\
&&+V_{3}[\sin \kappa _{i}\cos \kappa _{j}\cos 2(\psi _{0}\!-\!\psi
_{i})+\sin \kappa _{j}\cos \kappa _{i}\cos 2(\psi _{0}\!-\!\bar{\psi})]\Bigr
\}\ ,  \label{kappalossSS}
\end{eqnarray}
\ where $j\neq i$ and the coefficients $V_{1-3}$ are:
\begin{eqnarray}
V_{1} &=&12E^{2}\bar{L}^{4}+60G^{2}m^{2}\mu ^{3}E\bar{L}^{2}+35G^{4}m^{4}\mu
^{6}  \nonumber \\
V_{2} &=&564E^{2}\bar{L}^{4}+1620G^{2}m^{2}\mu ^{3}E\bar{L}
^{2}+805G^{4}m^{4}\mu ^{6}  \nonumber \\
V_{3} &=&60(8E^{2}\bar{L}^{4}+18G^{2}m^{2}\mu ^{3}E\bar{L}
^{2}+7G^{4}m^{4}\mu ^{6})\ .
\end{eqnarray}
Note that $V_{1}=D_{1}$, because both arose from the average of $r^{-6}.$
(The coefficient $D_{1}$, given by Eq. (4.32) of \cite{spinspin}, governs
the self-interaction term in the radiative loss of $L$). In order the $3/2$
post-Newtonian order-accurate spin-orbit contribution to hold at $2PN$, the
replacements $L\rightarrow \bar{L}$ and $A_{0}\rightarrow \bar{A}$ should be
carried on in Eq. (4.4)$\ $of \cite{GPV3} .
 
We stress that Eqs. (\ref{gammaloss})-(\ref{kappalossSS}) contain {\it all
radiative terms in the angular evolution up to the second post-Newtonian
order above the leading radiative effects. }
 
\section{Concluding remarks}
 
By proving the $5/2$ post-Newtonian accurate result that axisymmetric
objects do not radiate away any fraction of their initial spins, the
computation of the secular angular evolutions induced by radiation reaction
became simpler. Then from the analysis of $d{\bf J}/dt$ we could conclude
that there are no $N,\ PN,\ 2PN$ and tail contributions to the secular
radiative angular evolutions. The $SO$ part of the equations was given in
\cite{GPV3}. We have derived the $SS$ terms here.
 
No spin-spin terms appear in the radiative evolution of the angle $\gamma .$
As the spin-orbit terms average out \cite{GPV3}, we found the remarkable
result that the angle spanned by the spins receives no radiative secular
change up to the second post-Newtonian order. (However the angle $\gamma ,$
together with all other angles is subjected to {\it precessional} - both
instantaneous and secular - evolution. Although the second post-Newtonian 
order precessional (nonradiative) evolution of the angles is not developed 
in the paper, inspection of Table 1 shows that the precessions 
do not contribute to the 2PN radiative evolution of the angles $\kappa_i$ 
and $\gamma$, which contain only 3/2 PN and 2PN parts.)
 
The angles $\kappa _{i}$ evolve under the influence of both spin-orbit and
spin-spin terms. The radiative angular evolution equations
(\ref{gammaloss})-(\ref{kappalossSS})
together with the algebraic constraints
(\ref{ang1})-(\ref{ang2})
and the expressions for $dE/dt$ and $dL/dt$ derived in \cite
{spinspin} form a closed system of first order differential equations.
 
The angle $\psi _{0}$ appearing in this system is an integration constant.
It is interpreted as the angle subtended by the node line with ${\bf r}$ at $
\chi =0$ (with the periastron line). Each precession modifies the value of $
\psi _{0}$ by a small amount. According to the arguments of Ryan in \cite
{Ry2}, where this angle first appeared, the terms containing periodic
functions of $\psi _{0}$ average to zero whenever the precession time scale
is short compared to the radiation reaction time scale.
 
As it happened with the loss of $E$ and $L$ \cite{spinspin}, the spin-spin
terms of the radiative $\kappa _{i}$-evolution could be decomposed into
two-body and self-interaction terms. In the one-spin limit ${\bf S_{2}}=0$ the
terms proportional to $S_{1}^{2}$ from Eq.(\ref{kappalossself}) represent
the second post-Newtonian correction to the radiative evolution of the angle
$\kappa _{1}$ derived earlier in the Lense-Thirring approximation \cite{GPV1}.
 
\section{Acknowledgments}
 
This work has been supported by the Hungarian Scholarship Board. The
algebraic package REDUCE was employed in some of the computations.

\end{document}